\documentclass[twocolumn,trackchanges]{aastex63}
\usepackage{graphicx}	
\usepackage{amsmath}	
\usepackage{amssymb}	
\usepackage{float}
\usepackage{here}

\received{}
\revised{}
\accepted{}
\submitjournal{ApJL}

\shorttitle{The First TESS Self-Lensing Pulses}
\shortauthors{Sorabella et al.}
\graphicspath{{./}{figures/}}

\begin{document}

\title{The First TESS Self-Lensing Pulses: Revisiting KIC 12254688}

\correspondingauthor{Nicholas M. Sorabella}
\email{sorabella91@gmail.com}

\author[0000-0002-3562-9699]{Nicholas M. Sorabella}
\affiliation{Lowell Center for Space Science and Technology, Lowell, MA 01854}
\affiliation{Department of Physics and Applied Physics, University of Massachusetts Lowell, Lowell, MA 01854}

\author[0000-0002-8427-0766]{Silas G.T. Laycock}
\affiliation{Lowell Center for Space Science and Technology, Lowell, MA 01854}
\affiliation{Department of Physics and Applied Physics, University of Massachusetts Lowell, Lowell, MA 01854}

\author[0000-0002-7652-2206]{Dimitris M. Christodoulou}
\affiliation{Lowell Center for Space Science and Technology, Lowell, MA 01854}

\author[0000-0001-8572-8241]{Sayantan Bhattacharya}
\affiliation{Lowell Center for Space Science and Technology, Lowell, MA 01854}
\affiliation{Department of Physics and Applied Physics, University of Massachusetts Lowell, Lowell, MA 01854}



\begin{abstract}
We report the observations of two self-lensing pulses from KIC 12254688 in Transiting Exoplanet Survey Satellite (TESS) light curves.  This system, containing a F2V star and white-dwarf companion, was amongst the first self-lensing binary systems discovered by the Kepler Space Telescope over the past decade.  Each observed pulse occurs when the white dwarf transits in front of its companion star, gravitationally lensing the star's surface, thus making it appear brighter to a distant observer.  These two pulses are the very first self-lensing events discovered in TESS observations.  We describe the methods by which the data were acquired and detrended, as well as the best-fit binary parameters deduced from our self-lensing+radial velocity model.  We highlight the difficulties of finding new self-lensing systems with TESS, and we discuss the types of self-lensing systems that TESS may be more likely to discover in the future.

\end{abstract}



\section{Introduction}
Self-lensing binary systems are a subset of eclipsing binary systems in which at least one component is a compact object, such as a white dwarf (WD), neutron star (NS), or black hole (BH).  When the compact object transits in front of its (typically noncompact) companion, it gravitationally lenses the companion's surface, resulting in a symmetric flare in the system's light curve \citep{1973A&A....26..215M,1994ApJ...430..505W,2001MNRAS.324..547M,2011MNRAS.410..912R,2022ApJ...936...63S}.  Self-lensing pulses can be very useful in estimating binary parameters, as the pulse profile depends on the masses of the compact object and star, their separation during the transit, and the orbital inclination of the system.  Given that these pulses will occur every time the compact object transits its companion, self-lensing pulses are periodic, thus, they provide a way of also measuring the orbital period of the binary without the need for spectroscopic measurements.  The profile of the lensing pulse can also provide information regarding the star being lensed, such as its radius and limb-darkening coefficient. However, studies of self-lensing systems remain scarce, as only five stellar-mass self-lensing systems (and a sixth unconfirmed candidate) have been discovered to date \citep{Kruse275,2018AJ....155..144K,2019ApJ...881L...3M}.  All of these systems were discovered with the Kepler Space Telescope and all were found to be quite similar; each binary system contains a WD that gravitationally lenses a main-sequence companion not too dissimilar to that of our Sun.  

As a spiritual successor to the Kepler Space Telescope, the Transiting Exoplanet Survey Satellite (TESS) \citep{2010AAS...21545006R} is a very important tool in the search for new self-lensing pulses.  While Kepler had the advantage of obtaining more data for each target, as it was always observing the same patch of sky for 4 years, TESS has the distinct advantage of searching far more area of the sky.  TESS images a $24^{\circ} \times 96^{\circ}$ sector of the sky for 27 days before moving on to another sector. This drastically increases the number of stars observed and the potential for discovery of more self-lensing systems (as well as extrasolar systems).  It has been predicted by \citet{2021MNRAS.507..374W} that TESS could find anywhere from 60-170 self-lensing sources throughout its operational lifetime, increasing the number of known self-lensing systems by a factor of 10-30.  Combined with other surveys such as the Zwicky Transient Facility (ZTF) and the upcoming Legacy Survey of Space and Time (LSST), the number of detectable self-lensing systems could be of the same order as the population of X-ray binaries in the Milky Way.

We have begun our own search for new self-lensing pulses in the existing TESS data, although we have yet to find any new systems. Instead, we have found an abundance of asteroids crossing in front of background stars, giving false-positive signals that mimic self-lensing pulses \cite{2023ApJ...954...59S}. We also checked whether TESS has observed any of the known self-lensing WD systems \citep[][hereafter KMM18]{2018AJ....155..144K} at times when new pulses were expected to occur. This search highlighted a disadvantage of TESS observations.  As TESS observes a sector for only 27 days, if a binary were to have a much longer orbital period, there would be a much lower probability of observing it at the right time.  The four KMM18 self-lensing systems have orbital periods of $>$\,400 days, and three of them were not observed at times when their pulses were anticipated. One system (KIC 12254688), however, was not only observed at the right time, but rather fortuitously, TESS detected two pulses in Sectors 41 and 56, respectively.  These pulses mark the first two self-lensing events reported from TESS archival data, despite the fact that KIC 12254688 exhibits by far the smallest of the four self-lensing flares in the Kepler data.  

Furthermore, after the end of the primary mission of TESS, the longest-exposure Full Frame Images (FFIs) have cadences of 10 minutes compared to the 30-minute cadence of both the TESS primary mission and Kepler.  In the cases of Sectors 41 and 56, the cadence was even lower (see Section \ref{sec:2}), implying a signal-to-noise ratio (SNR) much lower than that achieved in the original Kepler detection of the KIC 12254688 pulse (KMM18).  While the low SNR complicated the extraction of the TESS pulses, we were still able to identify them and measure them in the noisy light curves of both sectors.

In Section \ref{sec:2}, we detail the methods by which we acquired, detrended, binned, and folded the two light curves from Sectors 41 and 56.  In Section \ref{sec:3}, we describe the methodology and the results of fitting our self-lensing model to the folded TESS pulse of KIC 12254688 before comparing our best-fit model parameters to those obtained by KMM18 and to those our model produced from the original Kepler data set. In Section \ref{sec:4}, we discuss the implications of this type of modeling, including the difficulties in discovering such lensing events in TESS data. In Section \ref{sec:5}, we summarize this work and our conclusions.

\section{Acquiring and Detrending the Pulses}
\label{sec:2}
Light curves for Sectors 41 and 56 were acquired using the Python-based time series analysis package {\tt Lightkurve} \citep{2018ascl.soft12013L}.  Specifically, we used {\tt Lightkurve}'s integrated TESScut \citep{2019ascl.soft05007B} function {\tt search\_tesscut} to get the data products for both sectors.  The individual frame exposure times for Sectors 41 and 56 were $475$ and $158$ seconds, respectively.  

We used different methods to remove TESS systematics from the two light curves.  For sector 41, we used {\tt Lightkurve}'s integrated Cotrending Basis Vectors function known as {\tt CBVCorrector}, which removes systematic noise.  Due to unknown errors from {\tt CBVCorrector} when applied to Sector 56, we relied instead on {\tt Lightkurve}'s {\tt RegressionCorrector} function in this sector.  {\tt RegressionCorrector} removes scattered light background and spacecraft motion from TESS light curves using linear regression with a design matrix constructed from the regressors.  

\begin{figure} 
    \centering
    \includegraphics[width=\columnwidth]{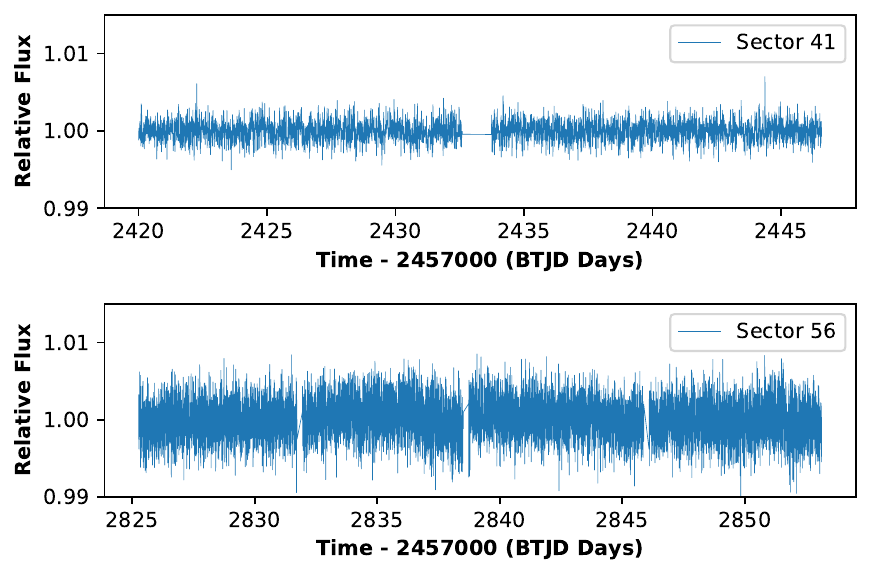}
        \caption{Detrended light curves of KIC 12254688 from TESS Sectors 41 and 56.  The projected pulse midpoints are expected to lie at 2424.734 BTJD and 2843.449 BTJD in Sectors 41 and 56, respectively.}
    \label{fig:1}
\end{figure}

The detrended light curves are shown in Figure \ref{fig:1}.  
One may notice that the pulses are not inherently visible in these detrended light curves (particularly in Sector 56), as the expected magnification from the lensing events (about 1.0005 in relative flux) is completely washed out by the noise intrinsic to each light curve.

\begin{figure} 
    \centering
    \includegraphics[width=\columnwidth]{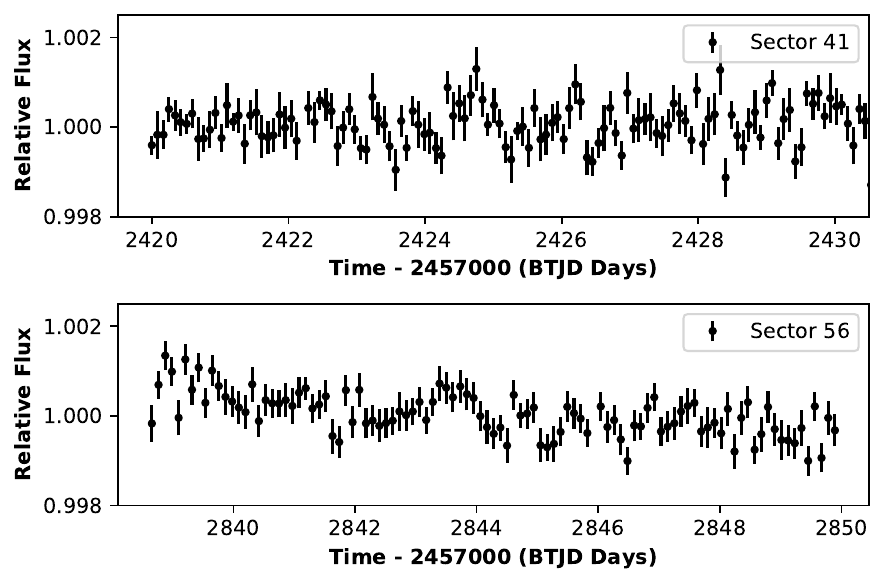}
        \caption{Binned light curves of KIC 12254688 for TESS Sectors 41 and 56, respectively.  We have zoomed in to the locations of the pulses (near the middle of each plot) so that they are more easily noticeable.  The midpoint of the Sector 41 pulse is at 2424.734 BTJD and Sector 56 is at 2843.449 BTJD.  Sector 56 has a small downward trend to the right of the pulse that we subsequently remove using a second-order polynomial fit, in which we exclude the pulse itself.}
    \label{fig:2}
\end{figure}

In hopes of extracting the pulses from these light curves, we binned each data set with the number of bins chosen to best reveal the pulses. For Sector 41, this ended up being 11 data points per bin, while for Sector 56, we assigned 45 points per bin.  The error bars were calculated using the error of the mean in each bin.  The resulting light curves are shown in Figure \ref{fig:2}.  At times 2424.734 BTJD and 2843.449 BTJD pulse shape profiles now appear in the data. We note that we used only the second half of Sector 56 data, in which the pulse was projected to appear.  A downward trend in seen throughout this light curve.  As this trend is not explained by self-lensing or any other relevant effect (e.g., Doppler-boosting kinematics or ellipsoidal variations due to tidal forces), we removed it by applying a second-order polynomial fit to the data on the sides of the pulse, after excluding the pulse itself.  This flattened the light curve on the sides, leaving only the pulse shape centered at 2843.449 BTJD to rise above the flat background.  

We normalized the light curves about the means of their respective data sets, excluding the pulses in both sets.  We then folded the two light curves using the orbital period from KMM18 ($418.715$ days), combining the two pulses into one better-sampled pulse.  Still, this combined TESS pulse shows much more variation than the original Kepler pulse due to the worse SNR in the TESS light curves.  This fact inexorably leads to the question of whether a self-lensing model fitted to these TESS data would return best-fit parameters similar to those determined by KMM18 from the Kepler light curve of KIC 12254688.   

\section{Fitting the Folded Pulse}
\label{sec:3}

As described in \citet{2022ApJ...927..234S}, we use our post-Newtonian orbital motion code derived from the Paczynski-Wiita potential to find the velocities and binary separation of the two stars in time. In this modeling effort, we exclude additional effects analyzed in \citet{2022ApJ...936...63S}, such as relativistic Doppler boosting and ellipsoidal variations.  While these effects are expected in some stellar-compact binary systems, they become significant only in systems with shorter orbital periods (smaller separations) than that of KIC 1225488 \citep{2019ApJ...883..169M}.

\subsection{Basic Equations}
The magnification caused by a self-lensing event \citep[e.g.,][]{Paczynski1986,1994ApJ...430..505W,2001MNRAS.324..547M,Lee2009,DD2018} depends mainly on (a) $\rho_{\star}(t)$, the time-dependent ratio between the radius of the star and the Einstein radius of the lens $R_{\rm E}(t)$, and (b) $u(t)$, the time-dependent projected separation between the lens and the source, expressed in units of the Einstein radius. The ratio $\rho_{\star}$ is given by 
\begin{equation}
    \rho_{\star}(t) = {\frac{R_{\star}}{R_{\rm E}(t)}}\, ,
    \label{eq1}
\end{equation}
where $R_{\star}$ is the radius of the source, and the Einstein radius is given by \begin{equation}
   R_{\rm E}(t)=\sqrt{\frac{4GM}{c^2} d(t) \sin(i)\,}\, ,
\end{equation}
where $G$ is the gravitational constant, $M$ is the mass of the compact object, $c$ is the speed of light, $d(t)$ is the time-dependent separation between the binary components, and $i$ is the inclination of the binary orbit to the plane of the sky (note that $i\approx 90^\circ$ for self-lensing binary systems).

The projected separation in units of $R_{\rm E}$ is determined as a function of time \citep{2011MNRAS.410..912R} by
\begin{equation}
    u(t)={\sqrt{u_{0}^2+\left({{\frac{t-t_0}{{t_{\rm E}}}}}\right)^2}}\, ,
    \label{eq_te}
\end{equation}
where $t_0$ is the time of closest projected approach and
\begin{equation}
\label{eq4}
    u_{\rm 0}=\frac{d(t_0)}{R_{\rm E}(t_0)} \sin{\psi} \, 
\end{equation}
is the impact parameter in units of $R_{\rm E}$ \citep{2021MNRAS.507..374W} at $t=t_0$, and $\psi$ is the inclination of the source relative to the observer-lens line-of-sight (so the orbital inclination is $i=90^\circ - \psi$).  The Einstein time $t_{\rm E}$ represents the time it takes for the lens to move across the length of its Einstein radius $R_{\rm E}$.

Finally, the time-dependent magnification $A(t)$ is expressed in terms of $u(t)$ and $\rho_{\star}$ by the equation \citep{2015AdAst2015E..11H}

\begin{equation}
   A=\frac{\displaystyle\int_{0}^{2\pi}\int_{0}^{\rho_{\star}}\left(\frac{\xi^2 +  2}{\xi\,\sqrt{\xi^2 + 4}}\right) S\,rdrd\theta }{2\pi\displaystyle\int_{0}^{\rho_{\star}}S\,rdr }\, ,
   \label{eq9}
\end{equation}
where polar coordinates $(r, \theta)$ are set on the projected surface of the star, and
\begin{equation}
\xi^2 = u^2+r^2-2ur\cos{\theta}\, .
\end{equation}
The function $S(r)$ represents the brightness profile of the source, for which we adopted the linear limb-darkening relation of \citet{2015AdAst2015E..11H} and \citet{Han_2016}, viz.
\begin{equation}
    S(r)=1-\Gamma\left[1-\frac{3}{2}\sqrt{1-\left(\frac{r}{\rho_{\star}}\right)^2}\, \right]\, ,
    \label{eq10}
\end{equation}
where $\Gamma$ is the linear limb-darkening coefficient.

\subsection{EMCEE Modeling}
To fit our model to the data set, we use EMCEE \citep{2013PASP..125..306F}, the {\it Python} implementation of the Affine Invariant Markov chain Monte Carlo Ensemble sampler \citep{2010CAMCS...5...65G}. The EMCEE code accepts an input parameter set $\{x_i\}$ and maximizes the log-likelihood function
\begin{equation}
\label{likelihood}
\ln P(y|\{x_i\}, \sigma_n)=-\frac{1}{2}\sum_n\left[\frac{(y_n-m_n)^2}{\sigma_n^{~2}}+\ln \sigma_n^{~2}\right] ,
\end{equation}
where $y_n$ are the light-curve data points, $\sigma_n$ are the summations of the photometric error at each point and additional jitter in the data that the model doesn't account for ($f_{LC}$), and $m_n$ are the model's points given set $\{x_i\}$.  This is accomplished by the {\it Python} module {\tt scipy.optimize} \citep{python1995} which returns the most likely $\{x_i\}$ values, but not their uncertainties. These are obtained from the posterior probability distribution which is derived from $P(y|\{x_i\}, \sigma_n)$ and a predetermined prior function $P(m_n)$, viz.
\begin{equation}
P(m_n|\{x_i\},y_n,\sigma_n) \propto P(m_n)\,P(y|\{x_i\}, \sigma_n)\, .
\end{equation}  
The prior allows one to encode any prior information concerning the free parameters, and thus to preclude nonphysical results.

\begin{figure*} 
    \centering
    \includegraphics[width=6.5in]{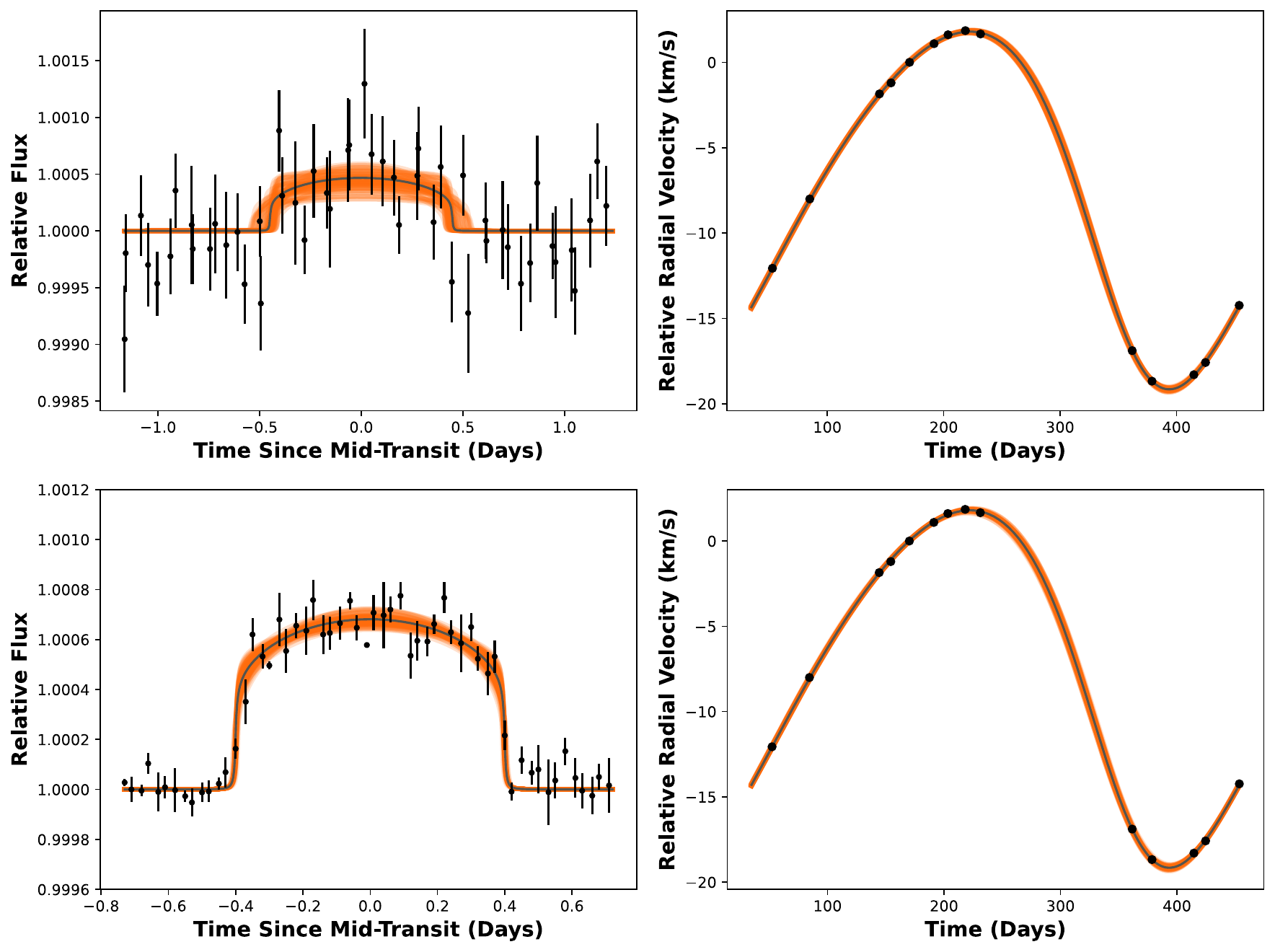}
        \caption{Our joint fits to the phase folded light curves of KIC 12254688 from TESS (top left) and Kepler (bottom left) with updated radial velocity data (right panels) from \citet{2020IAUS..357..215M}.  The orange lines represent the 95\% confidence intervals, and the dark gray lines represent the $50^{\rm th}$ percentile estimates.}
    \label{fig:3}
\end{figure*}

In order to compare our results with those obtained by KMM18, we also fit our model to the phase-folded self-lensing pulses originally observed by the Kepler Space Telescope.  As fitting to self-lensing pulses on their own can prove difficult given the degeneracies between parameters such as white dwarf mass and stellar radius, we decided to do a joint fit of the self-lensing data and radial velocity data of the system in similar fashion to KMM18.  The radial velocity (RV) data is as reported in \citet{2020IAUS..357..215M}, which was collected from the Tillinghast Reflector Echelle Spectrograph (TRES), while the RV model we use is calculated from the projected 2D velocity components of the star as found by our post-Newtonian orbital motion code.  This data set is an updated version of the one found in KMM18 and covers nearly a full orbit of the system.  It is also worth noting that these are relative radial velocities, where the motion of the binary system's barycenter is compared relative to another in the sample (whose RV is set to zero).  As such, we included a radial velocity offset term $RV_0$, to shift the RV curve to account for this relative motion. 

As the initial conditions of our orbital motion code set the binary components at apoapsis, an additional parameter we dubbed the time offset ($t_i$) is included in our modeling to shift their orbital position in time to match the starting point of the radial velocity and light curve data. This joint fit method was done for both the Kepler and TESS light curves and was accomplished by maximizing the summation of the likelihood functions of the self-lensing and radial velocity models. As it is a combination of two models, we added an additional jitter term, specific to the radial velocity model ($f_{RV}$).  

\subsection{Priors}\label{prior}
In the self-lensing models of KIC 12254688, the prior function allows us to constrain EMCEE by the known properties of the free parameters adopted in this study: (a) ranges for the mass and the radius of the primary star based on spectroscopic results \citep[KIC DR25,][]{Mathur2017}, (b) a plausible range for the mass of the white dwarf companion \citep[Figures 1, 3, 6 in][]{Kepler2007}, (c) a range for the orbital inclination \citep[Figure 8 in][]{2022ApJ...936...63S}, and (d) the range for the limb-darkening coefficient, which is set to (0, 1) by its definition.  The priors for the RV model are based closely on those used by KMM18 while the parameters not found in the KMM18 (or were derived parameters) were given, within in reason, wide priors.

\subsection{Best-Fit Models}
Our model fits to the TESS and Kepler pulses and their corresponding RV fits are show in Figure \ref{fig:3}, while the best-fit parameters are listed in Table \ref{tab:1}.  The corresponding corner plots of the posterior distributions are shown in Appendix~\ref{appa}.  Following standard practices, parameter values and their 1$\sigma$ errors were obtained from the 16, 50, and 84 percentiles of the posterior distributions.  The thickness of each orange curve in Figure \ref{fig:3} represents the $95\%$ confidence interval, as this was determined from $95\%$ of the walker chains chosen randomly from the set of all walkers. The $50^{th}$ percentile estimates are shown as the dark gray line.

While our model possesses some differences with that of the KMM18 model, our best-fit parameter values from the Kepler data set are, in general, very consistent with the original results of KMM18 (last column in Table~\ref{tab:1}).  The parameters that showed a significant difference ($>1\sigma$) compared with the KMM18 results were that of the radial velocity offset and the light curve jitter.    The difference in the RV offset is undoubtedly a consequence of using the updated data from \citet{2020IAUS..357..215M}, which reported $RV_0=7.49_{-0.3}^{+0.2}$, very much in line with our estimation.  As for the light curve jitter, this could easily stem from a difference in methodology in phase folding and binning the Kepler data.  In both cases, the estimated jitter in the light curve is very small (about a $0.01\%$ relative flux).

\begin{table*}
\caption{Best-fit self-lensing+RV models for KIC 12254688 utilizing the listed priors and the final 4000 steps of 100 independent walker chains in each data set. The returned values are obtained from the 16, 50 and 84 percentiles of their posterior distributions. The best-fit values from Table 4 of \citet{2018AJ....155..144K} (KMM18) are also listed in the last column for comparison.}
    \label{tab:1}
    \centering
    \begin{tabular}{lrcrcrl}
    \hline
                           & TESS Data & Adopted & Kepler Data    && Kepler Data \\ 
     Parameter Name (Unit) & Our Best Fit & Prior & Our Best Fit && KMM18 Fit   \\
     \hline
     \noalign{\vskip 0.75mm}
     White Dwarf Mass ($M_\odot$) & $0.60^{+0.04}_{-0.05}$ & (0.1,\,1.44) & $0.61^{+0.03}_{-0.04}$ && $0.62^{+0.09}_{-0.06}$ \\
     \noalign{\vskip 0.75mm}
     Primary Star Mass ($M_\odot$) & $1.52^{+0.19}_{-0.21}$ & (1.2,\,1.8) & $1.57^{+0.15}_{-0.17}$ && $1.5^{+0.1}_{-0.1}$ \\
     \noalign{\vskip 0.75mm}
     Stellar Radius ($R_\odot$) & $2.57^{+0.32}_{-0.25}$ & (1.5,\,3.5) & $2.16^{+0.08}_{-0.08}$ && $2.2^{+0.2}_{-0.1}$ \\
     \noalign{\vskip 0.75mm}
     Orbit Inclination (deg) & $89.78^{+0.13}_{-0.12}$ & (85,\,90) & $89.94^{+0.04}_{-0.04}$ && $\left[89.95_{-0.04}^{+0.05}\right]$ & \!\!\!\!\!\!$^{(a)}$ \\
     \noalign{\vskip 0.75mm}
     Limb Darkening Coefficient &  $0.50^{+0.33}_{-0.33}$ & (0,\,1) & $0.39^{+0.11}_{-0.09}$ && $\left[0.50_{-0.35}^{+0.34}\right]$ & \!\!\!\!\!\!$^{(b)}$ \\
     \noalign{\vskip 0.75mm}
     Eccentricity &  $0.19^{+0.01}_{-0.01}$ & (0,\,1) & $0.18^{+0.01}_{-0.01}$ && $0.20^{+0.04}_{-0.04}$ \\
     \noalign{\vskip 0.75mm}
     Argument of Periapsis (deg) &  $129.27^{+1.26}_{-1.34}$ & (100,\,150) & $129.29^{+1.29}_{-1.39}$ && $129^{+2.00}_{-2.00}$ \\
     \noalign{\vskip 0.75mm}
     Radial Velocity Offset, RV$_0$ (km s$^{-1}$) &  $-7.45^{+0.05}_{-0.05}$ & ($-15,\,15$) & $-7.45^{+0.05}_{-0.05}$ && $-7.7^{+0.4}_{-0.5}$ \\
     \noalign{\vskip 0.75mm}
     Time Offset, $T_i$ (days) &  $175.54^{+1.69}_{-1.60}$ & (100,\,300) & $175.97^{+1.78}_{-1.63}$ && $\cdots$ \\
     \noalign{\vskip 0.75mm}
     Light Curve Jitter, $\log(f_{LC})$ &  $-4.93^{+0.74}_{-0.72}$ & ($-6,\,-3$) & $-4.36^{+0.08}_{-0.08}$ && $-3.96^{+0.04}_{-0.04}$ \\
     RV Jitter, $\log(f_{RV})$ &  $-2.63^{+0.31}_{-0.25}$ & ($-3,\,3$) & $-2.63^{+0.30}_{-0.25}$ && $-1.92^{+0.70}_{-0.78}$ \\
     \noalign{\vskip 0.75mm}
     \noalign{\vskip 0.75mm}
     \hline
    \end{tabular}
    \vskip 0.75mm
    ${(a)}$~Inclination $i$ calculated from equation\,(\ref{bratio}). 
    ${(b)}$~Linear coefficient $u_1$ calculated from the $q$-coefficients in Table 4 of KMM18 ($q_1=0.7^{+0.2}_{-0.3}$\,,~ $q_2=0.3\pm0.2$) using $u_1=2q_2\sqrt{q_1}$ \citep{Kipping2013} and error propagation.
\end{table*}

The results for the TESS fit were largely the same as the Kepler fit, though with greater uncertainties for most parameters.  However, there were a couple of parameter estimates that showed significant difference compared to the KMM18 results (light curve jitter and radial velocity offset) and two more that showed significant difference with both KMM18 and that of our own Kepler fit (orbital inclination and stellar radius).  The difference with the radial velocity offset certainly stems from the same reason as mentioned for our Kepler fit.  The light curve jitter term actually came out to be smaller than even the Kepler fit, but this is likely a result of the already much larger photometric error present in the TESS light curve.

The difference between the Kepler and TESS orbital inclination estimates is significant though both place the system as nearly edge-on as expected.  As for stellar radius, the estimated value from the TESS fit is much higher than both the KMM18 result and our own Kepler estimate, even to the point that the error bars did not overlap with either result.  It is also worth noting that the uncertainty for $\Gamma$ was much higher than in the Kepler fit as well.  Both the uncertainty in $\Gamma$ and the higher value of $R_{\star}$ are certainly consequences of the significantly less defined nature of the TESS pulse compared with its Kepler counterpart.  Without a definitive shape, it is nearly impossible to determine the limb-darkening coefficient which ends up driving the stellar radius to higher values.

While it is clear visually in this TESS data that there is a relative flux increase of similar magnitude and timescale to that of the Kepler pulses, the higher intrinsic variability of the TESS light curves makes parameter estimation more uncertain.  While not diverting too far off from the estimates made by both our and KMM18's fit to the Kepler data, the larger size of the error bars and less Gaussian shape of the posterior distributions (see Appendix~\ref{appa}) would inexorably lead to the conclusion that the estimates from the Kepler data are more reliable.  Furthermore, fitting to the RV data without joint fitting with the TESS data led to markedly similar results as to what we saw from our TESS joint fit.  This would mean much of the parameter determination with the TESS model stems from the quality of the RV data and not the light curve. This coupled with the fact that the parameters that are only determined by the self-lensing model (stellar radius and limb-darkening coefficient) show significant differences with either Kepler model, is a sign that the TESS pulse does not have the required signal-to-noise to accurately estimate the binary parameters.  Regardless, the presence of TESS pulses at exactly the predicted times, with similar timescale and height to that of the original Kepler events, provides confidence that these are indeed the anticipated self-lensing pulses, making these signals the first self-lensing events reported from TESS observations.

\section{Variability, Synthetic Light Curves, and Discussion}
\label{sec:4}
\begin{figure*} 
    \centering
    \includegraphics[width=6.5in]{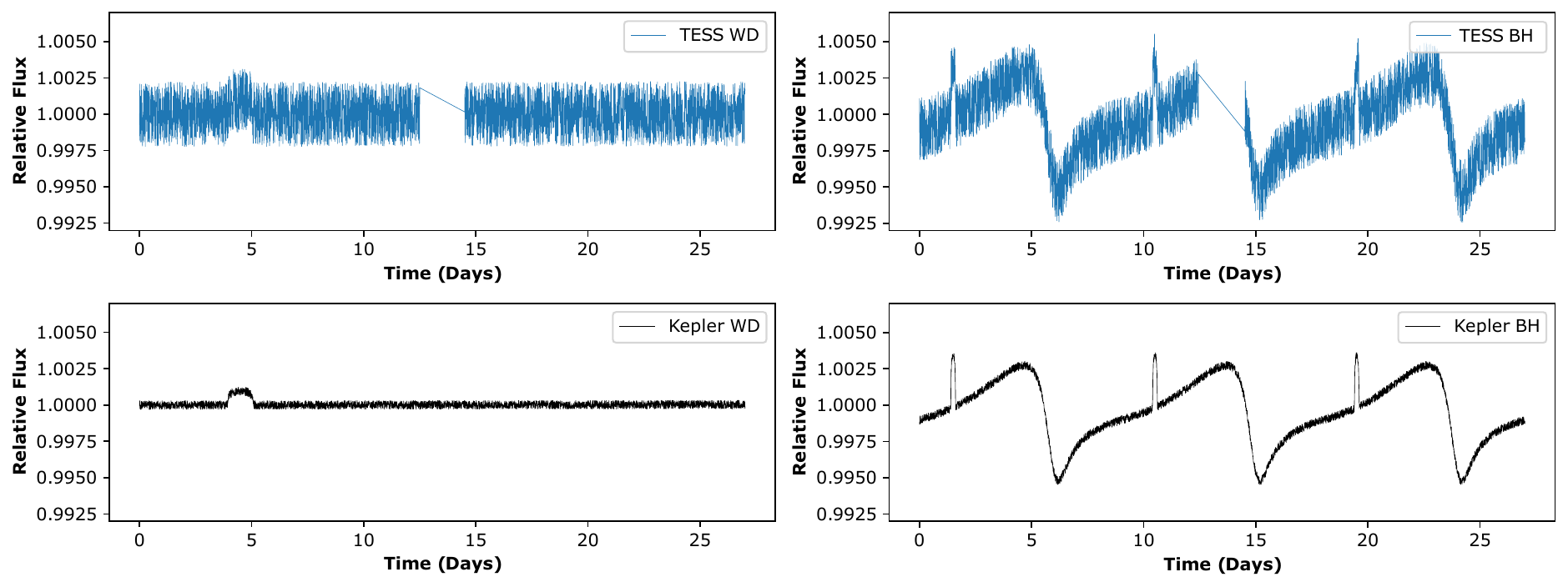}
        \caption{Synthetic light curves of a self-lensing binary system containing a $1.1 M_\odot$ star with a radius of $1.5 R_\odot$ and a $0.35 M_\odot$ WD (left panels) or a $7 M_\odot$ BH (right panels), with orbital periods of $270$ and $9$ days, respectively.  The top panels simulate the light curves as viewed by the $475$-s  ($\sim8$ min) cadence {\tt TESScut} exposures, and the bottom panels are the same light curves as viewed by the 30-min cadence of Kepler.}
    \label{fig:4}
\end{figure*} 

It is important to note the challenges in detecting self-lensing pulses, such as those reported here, in TESS data, which is far noisier than Kepler at the same apparent magnitude. Exposure times for TESS have been shortened in the extended mission from 2020 through the present. This is a decision which may merit re-examination for future cycles to restore sensitivity to small-amplitude pulses. KIC 12254688 was a known system, so we knew exactly when each pulse would occur; even then, further analysis had to be done to extract the pulses from the light curves.  Had this been an unknown system, it would have been difficult to discern these pulses from intrinsic variability in the detrended light curves.  This highlights an important limitation for lensing in TESS data as compared to Kepler data.  In the Kepler observations where KMM18 detected lensing events, the pulses stuck out visually in the light curves. This may not be the case in TESS observations since the other known self-lensing WD systems exhibit only slightly higher pulses than KIC 12254688. Thus, despite the projections for many more self-lensing systems containing WDs rather than NSs or BHs \citep[e.g.,][]{2021MNRAS.507..374W}, their smaller magnifications make WD lenses harder to detect in TESS data sets.

White dwarf lensing events also have the disadvantage of requiring longer orbital periods (larger separations) to be discernible, so TESS is likely to see just one pulse for most targets. Even if such a pulse were observed clearly in a sector, it would still be difficult to confirm the detection as an actual lensing event. To underscore the difficulties with WD lenses in (sub)10-min cadence TESS data, we generated synthetic light curves for a variety of cases (Figure \ref{fig:4}).  In all four panels, the source is a $1.1 M_\odot$ star with radius $1.5 R_\odot$ and the same TESS magnitude as KIC 12254688 ($T=12.67$), in an eccentric orbit ($e=0.5$ and $\omega=270^{\circ}$).  In the left panels, the compact object is a $0.35 M_\odot$ WD with an orbital period of $270$ days.  The $\sim$\,0.2\% magnification due to self-lensing is about the average of all currently known WD lensing magnifications. In the right panels, the compact object is a $7 M_\odot$ BH with an orbital period of only 9 days.  The top panels show the signals recorded with a $475$-s cadence, the same cadence as for TESS Sector 41; the bottom panels show the same signals recorded with the SNR seen in the 30-min cadence of Kepler.

While the pulse sticks out in the TESS WD light curve, it is hardly as prominent as the Kepler WD pulse.  In the absence of another detection, it would be difficult to argue for the origin of the TESS pulse. Since the search for self-lensing pulses inevitably requires searching through thousands of light curves, it may be easy for researchers to overlook such signals without doing additional data processing. Additional processing requires more time, time which could be spent looking for obvious signals in other data sets. Furthermore, the data gaps in the middle of each TESS light curve last for days, further reducing the observing duty-cycle for each target.

In the BH panels of Figure \ref{fig:4}, there are three self-lensing pulses, each with a much higher magnification ($\sim$\,0.4\%), due to the higher mass of the compact object.  Given the much shorter orbital period, additional physical effects, such as Doppler boosting and ellipsoidal variations, appear in the light curves.  Including these effects in the models further constrains the free parameters.  For instance, the new effects are highly dependent on the eccentricity of the orbit, a parameter that the pure lensing model cannot model on its own with great certainty. Thus, for BH lenses the orbital parameters can be tightly constrained even without follow-up spectroscopic measurements, in contrast to our present analysis of the KMM18 WD lens.  BH lenses are expected to be extremely rare, but if detected their light curves will provide unambiguous self-lensing signals despite the increased noise in TESS photometry relative to Kepler.

\section{Summary and Conclusions}\label{sec:5}

We have reported observations of the first two self-lensing pulses by the TESS Space Telescope.  These pulses occurred in the known self-lensing binary KIC 12254688, where a WD gravitationally lenses its stellar companion once per orbit.  We detrended the light curves from TESS Sectors 41 and 56 (Figure~\ref{fig:1}), binned the data (Figure~\ref{fig:2}), and folded them to provide a better sampled pulse.  We joint fitted our self-lensing and radial velocity models to both the folded TESS pulse and the original folded Kepler pulse and corresponding radial velocity data from \citet{2020IAUS..357..215M} (Figure~\ref{fig:3}).  We obtained best-fit parameters (Table \ref{tab:1}) that were, overall, consistent with those determined by the original study of \cite{2018AJ....155..144K} of the Kepler data.  However, we found inconsistencies in some of the returned TESS parameter estimates specific to the self-lensing model that are likely caused by the more chaotic, less defined nature of the folded TESS pulse.  These parameters are still within reason, but are significantly different than what we and KMM18 found for the Kepler pulses.     

It has been suggested by \citet{2019ApJ...883..169M} and \citet{2021MNRAS.507..374W} that we are going to discover many more Galactic self-lensing systems containing WDs than NSs or BHs in TESS data and this could very well be the case.  On the other hand, we have highlighted the difficulty in parameter estimation in such lower magnification systems in the lower-cadence extended TESS mission as well as the difficulties that we will face in even distinguishing such WD self-lensing pulses from intrinsic variability (\S~\ref{sec:4}).  Thus, it may be wise to first search the higher-cadence TESS data obtained during the primary mission. We also note the possibility of detecting a BH lensing event by mere chance.  In such a case, it would be much easier to identify its repeated short-period pulses in the pipeline data, especially in the lower-cadence TESS data of the extended mission.

\begin{acknowledgments}

We would like to thank Kento Masuda and the members of the TRES team for providing the radial velocity measurements of KIC 12254688, and the members of the Lowell Center for Space Science and Technology (LoCSST) for their support throughout this investigation. Detailed reviewer comments and suggestions are also greatly appreciated. This work was supported by a NASA FINESST grant, NSF grant AST-2109004, and the Massachusetts Space Grant Consortium.

The data presented in this paper were obtained from the Mikulski Archive for Space Telescopes (MAST). The particular sets of observations that we analyzed can be downloaded from link\dataset[10.17909/hm87-b313]{\doi{10.17909/hm87-b313}}.

\software{EMCEE \citep{2010CAMCS...5...65G, 2013PASP..125..306F}, 
Python module {\tt corner.py} \citep{Foreman-Mackey2016,Hogg2018}, 
Python SciPy module {\tt optimize} \citep[\url{https://scipy.org},][]{python1995},
Python package {\tt Lightkurve} \citep{2018ascl.soft12013L}, 
Mathematica, v. 13.3 (\url{https://www.wolfram.com/mathematica}), 
MATLAB, v. R2023b (\url{https://www.mathworks.com}).
}
\end{acknowledgments}




\bibliography{self_lensing} 

\begin{thebibliography}{}
\expandafter\ifx\csname natexlab\endcsname\relax\def\natexlab#1{#1}\fi
\providecommand{\url}[1]{\href{#1}{#1}}
\providecommand{\dodoi}[1]{doi:~\href{http://doi.org/#1}{\nolinkurl{#1}}}
\providecommand{\doeprint}[1]{\href{http://ascl.net/#1}{\nolinkurl{http://ascl.net/#1}}}
\providecommand{\doarXiv}[1]{\href{https://arxiv.org/abs/#1}{\nolinkurl{https://arxiv.org/abs/#1}}}

\bibitem[{{Brasseur} {et~al.}(2019){Brasseur}, {Phillip}, {Fleming}, {et~al.}}]{2019ascl.soft05007B}
{Brasseur}, C.~E., {Phillip}, C., {Fleming}, S.~W., {et~al.} 2019, {Astrocut: Tools for creating cutouts of TESS images}, ASCL record ascl:1905.007.
\newblock \doeprint{1905.007}

\bibitem[{{Cardoso} {et~al.}(2018){Cardoso}, {Hedges}, {Gully-Santiago}, {Saunders}, {Cody}, {Barclay}, {Hall}, {Sagear}, {Turtelboom}, {Zhang}, {Tzanidakis}, {Mighell}, {Coughlin}, {Bell}, {Berta-Thompson}, {Williams}, {Dotson}, \& {Barentsen}}]{2018ascl.soft12013L}
{Cardoso}, J.~V.~d.~M., {Hedges}, C., {Gully-Santiago}, M., {et~al.} 2018, {Lightkurve: Kepler and TESS time series analysis in Python}, Astrophysics Source Code Library.
\newblock \doeprint{1812.013}

\bibitem[{{D'Orazio} \& {Di Stefano}(2018)}]{DD2018}
{D'Orazio}, D.-J., \& {Di Stefano}, R. 2018, \mnras, 474, 2975, \dodoi{10.1093/mnras/stx2936}

\bibitem[{Foreman-Mackey(2016)}]{Foreman-Mackey2016}
Foreman-Mackey, D. 2016, Journal of Open Source Software, 1, 24, \dodoi{10.21105/joss.00024}

\bibitem[{{Foreman-Mackey} {et~al.}(2013){Foreman-Mackey}, {Hogg}, {Lang}, \& {Goodman}}]{2013PASP..125..306F}
{Foreman-Mackey}, D., {Hogg}, D.~W., {Lang}, D., \& {Goodman}, J. 2013, \pasp, 125, 306, \dodoi{10.1086/670067}

\bibitem[{{Goodman} \& {Weare}(2010)}]{2010CAMCS...5...65G}
{Goodman}, J., \& {Weare}, J. 2010, Communications in Applied Mathematics and Computational Science, 5, 65, \dodoi{10.2140/camcos.2010.5.65}

\bibitem[{{Hamolli} {et~al.}(2015){Hamolli}, {Hafizi}, {De Paolis}, \& {Nucita}}]{2015AdAst2015E..11H}
{Hamolli}, L., {Hafizi}, M., {De Paolis}, F., \& {Nucita}, A.~A. 2015, Advances in Astronomy, 2015, 402303, \dodoi{10.1155/2015/402303}

\bibitem[{Han(2016)}]{Han_2016}
Han, C. 2016, ApJ, 820, 53, \dodoi{10.3847/0004-637x/820/1/53}

\bibitem[{{Hogg} \& {Foreman-Mackey}(2018)}]{Hogg2018}
{Hogg}, D.~W., \& {Foreman-Mackey}, D. 2018, \apjs, 236, 11, \dodoi{10.3847/1538-4365/aab76e}

\bibitem[{{Kawahara} {et~al.}(2018){Kawahara}, {Masuda}, {MacLeod}, {Latham}, {Bieryla}, \& {Benomar}}]{2018AJ....155..144K}
{Kawahara}, H., {Masuda}, K., {MacLeod}, M., {et~al.} 2018, \aj, 155, 144 (KMM18), \dodoi{10.3847/1538-3881/aaaaaf}

\bibitem[{{Kepler} {et~al.}(2007){Kepler}, Kleinman, Nitta, {et~al.}}]{Kepler2007}
{Kepler}, S.~O., Kleinman, S.~J., Nitta, A., {et~al.} 2007, \mnras, 375, 1315, \dodoi{10.1111/j.1365-2966.2006.11388.x}

\bibitem[{{Kipping}(2013)}]{Kipping2013}
{Kipping}, D.~M. 2013, \mnras, 435, 2152, \dodoi{10.1093/mnras/stt1435}

\bibitem[{Kruse \& Agol(2014)}]{Kruse275}
Kruse, E., \& Agol, E. 2014, Science, 344, 275, \dodoi{10.1126/science.1251999}

\bibitem[{{Lee} {et~al.}(2009){Lee}, Riffeser, Seitz, \& Bender}]{Lee2009}
{Lee}, C.-H., Riffeser, A., Seitz, S., \& Bender, R. 2009, \apj, 695, 200, \dodoi{10.1088/0004-637X/695/1/200}

\bibitem[{{Maeder}(1973)}]{1973A&A....26..215M}
{Maeder}, A. 1973, \aap, 26, 215

\bibitem[{{Marsh}(2001)}]{2001MNRAS.324..547M}
{Marsh}, T.~R. 2001, \mnras, 324, 547, \dodoi{10.1046/j.1365-8711.2001.04293.x}

\bibitem[{{Masuda} \& {Hotokezaka}(2019)}]{2019ApJ...883..169M}
{Masuda}, K., \& {Hotokezaka}, K. 2019, \apj, 883, 169, \dodoi{10.3847/1538-4357/ab3a4f}

\bibitem[{{Masuda} {et~al.}(2019){Masuda}, {Kawahara}, {Latham}, {Bieryla}, {Kunitomo}, {MacLeod}, \& {Aoki}}]{2019ApJ...881L...3M}
{Masuda}, K., {Kawahara}, H., {Latham}, D.~W., {et~al.} 2019, \apjl, 881, L3, \dodoi{10.3847/2041-8213/ab321b}

\bibitem[{{Masuda} {et~al.}(2020){Masuda}, {Kawahara}, {Latham}, {Bieryla}, {MacLeod}, {Kunitomo}, {Benomar}, \& {Aoki}}]{2020IAUS..357..215M}
{Masuda}, K., {Kawahara}, H., {Latham}, D.~W., {et~al.} 2020, in White Dwarfs as Probes of Fundamental Physics: Tracers of Planetary, Stellar and Galactic Evolution, ed. M.~A. {Barstow}, S.~J. {Kleinman}, J.~L. {Provencal}, \& L.~{Ferrario}, Vol. 357, 215--219, \dodoi{10.1017/S1743921320000915}

\bibitem[{{Mathur} {et~al.}(2017){Mathur}, Huber, Batalha, {et~al.}}]{Mathur2017}
{Mathur}, S., Huber, D., Batalha, N.~M., {et~al.} 2017, \apjs, 229, 30, \dodoi{10.3847/1538-4365/229/2/30}

\bibitem[{{Paczy\'nski}(1986)}]{Paczynski1986}
{Paczy\'nski}, B. 1986, \apj, 304, 1

\bibitem[{{Rahvar} {et~al.}(2011){Rahvar}, {Mehrabi}, \& {Dominik}}]{2011MNRAS.410..912R}
{Rahvar}, S., {Mehrabi}, A., \& {Dominik}, M. 2011, \mnras, 410, 912, \dodoi{10.1111/j.1365-2966.2010.17490.x}

\bibitem[{{Ricker} {et~al.}(2010){Ricker}, {Latham}, {Vanderspek}, {Ennico}, {Bakos}, {Brown}, {Burgasser}, {Charbonneau}, {Clampin}, {Deming}, {Doty}, {Dunham}, {Elliot}, {Holman}, {Ida}, {Jenkins}, {Jernigan}, {Kawai}, {Laughlin}, {Lissauer}, {Martel}, {Sasselov}, {Schingler}, {Seager}, {Torres}, {Udry}, {Villasenor}, {Winn}, \& {Worden}}]{2010AAS...21545006R}
{Ricker}, G.~R., {Latham}, D.~W., {Vanderspek}, R.~K., {et~al.} 2010, in American Astronomical Society Meeting Abstracts, Vol. 215, American Astronomical Society Meeting Abstracts \#215, 450.06

\bibitem[{{Sorabella} {et~al.}(2022{\natexlab{a}}){Sorabella}, {Bhattacharya}, {Laycock}, {Christodoulou}, \& {Massarotti}}]{2022ApJ...936...63S}
{Sorabella}, N.~M., {Bhattacharya}, S., {Laycock}, S. G.~T., {Christodoulou}, D.~M., \& {Massarotti}, A. 2022{\natexlab{a}}, \apj, 936, 63, \dodoi{10.3847/1538-4357/ac82b7}

\bibitem[{{Sorabella} {et~al.}(2022{\natexlab{b}}){Sorabella}, {Bhattacharya}, {Laycock}, {Christodoulou}, \& {Massarotti}}]{2022ApJ...927..234S}
---. 2022{\natexlab{b}}, \apj, 927, 234, \dodoi{10.3847/1538-4357/ac4a59}

\bibitem[{{Sorabella} {et~al.}(2023){Sorabella}, {Laycock}, {Neeley}, {Christodoulou}, \& {Bhattacharya}}]{2023ApJ...954...59S}
{Sorabella}, N.~M., {Laycock}, S. G.~T., {Neeley}, L.~J., {Christodoulou}, D.~M., \& {Bhattacharya}, S. 2023, \apj, 954, 59, \dodoi{10.3847/1538-4357/ace9df}

\bibitem[{Van~Rossum \& Drake(1995)}]{python1995}
Van~Rossum, G., \& Drake, Jr, F.~L. 1995, Python Reference Manual (CWI, Amsterdam).
\newblock \url{https://ir.cwi.nl/pub/5008}

\bibitem[{{Wiktorowicz} {et~al.}(2021){Wiktorowicz}, {Middleton}, {Khan}, {Ingram}, {Gandhi}, \& {Dickinson}}]{2021MNRAS.507..374W}
{Wiktorowicz}, G., {Middleton}, M., {Khan}, N., {et~al.} 2021, \mnras, 507, 374, \dodoi{10.1093/mnras/stab2135}

\bibitem[{{Witt} \& {Mao}(1994)}]{1994ApJ...430..505W}
{Witt}, H.~J., \& {Mao}, S. 1994, \apj, 430, 505, \dodoi{10.1086/174426}

\end{thebibliography}

\appendix
\restartappendixnumbering
\section{MCMC Statistics}\label{appa}

\subsection{Corner Plots}
\begin{figure}[b] 
    \centering
    \includegraphics[width=6in]{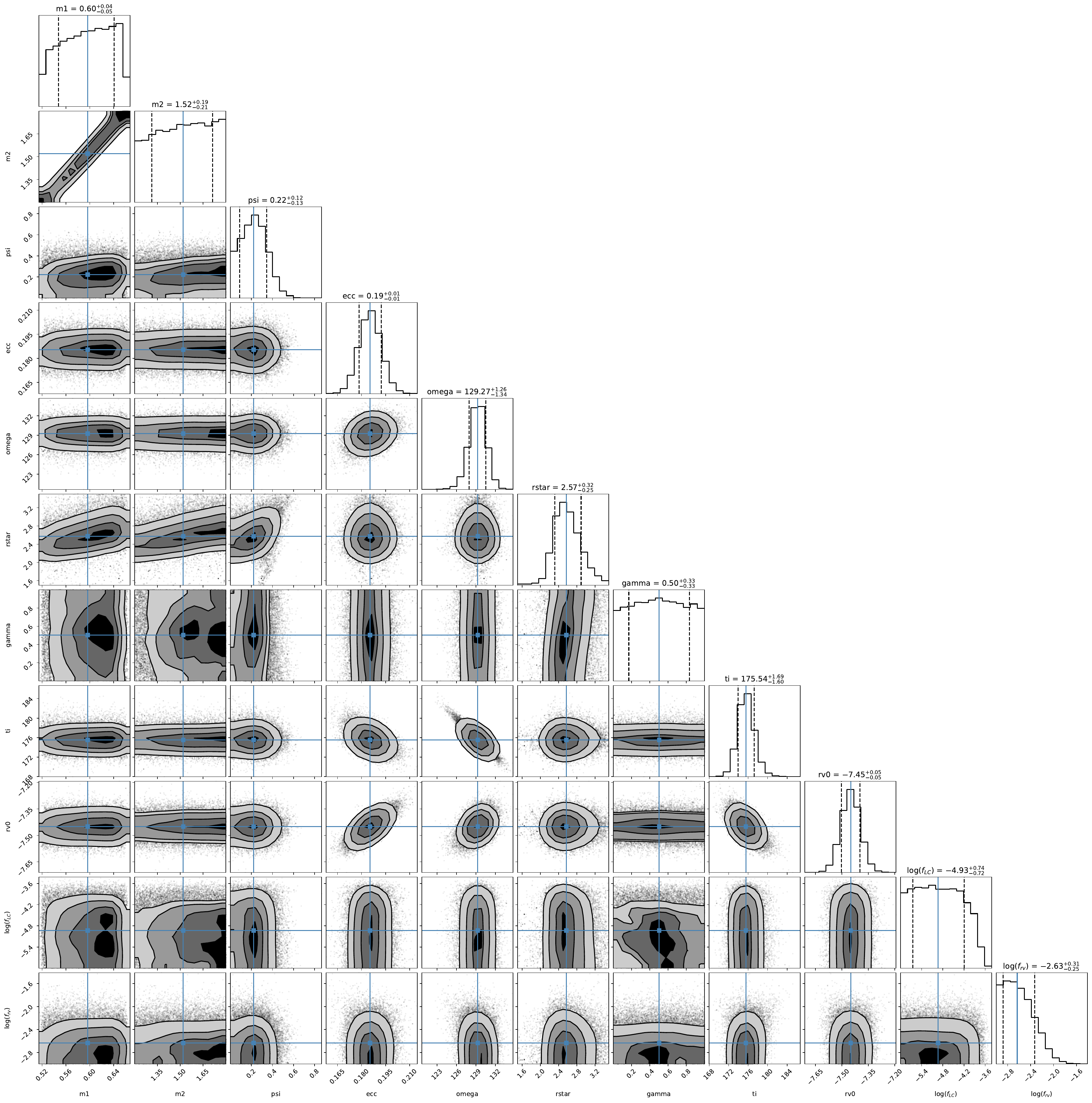}
        \caption{One and two dimensional projections of the posterior probability distributions of the model parameters for the TESS data fit. The returned values are obtained from the 16, 50 and 84 percentiles, utilizing the final 4000 steps of 100 independent MCMC walker chains.}
    \label{fig:5}
\end{figure}

\begin{figure} 
    \centering
    \includegraphics[width=6in]{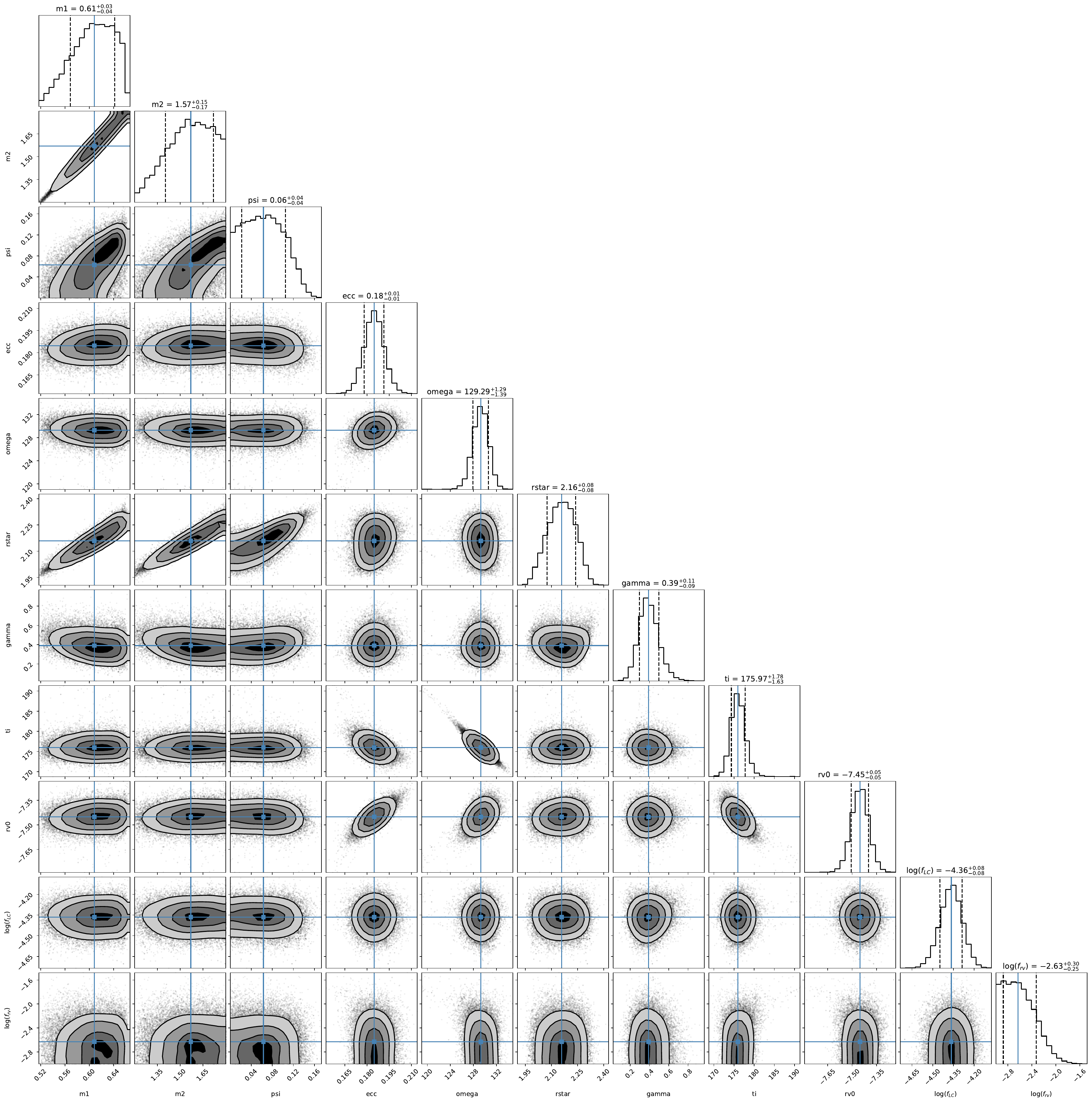}
        \caption{As in Figure~\ref{fig:5}, but for the Kepler data fit.}
    \label{fig:6}
\end{figure}

Corner plots were generated by the {\it Python} module {\tt corner.py} \citep{Foreman-Mackey2016,Hogg2018} from the derived posterior distributions of the free parameters of our models. Figures \ref{fig:5} and \ref{fig:6} display the corner plots from the TESS and the Kepler data fits, respectively.

In general, the posterior distributions of the TESS fit were much wider and less Gaussian than that of their Kepler counterparts.  The stellar mass (`m2' in columns 2) and the limb-darkening coefficient $\Gamma$ were not constrained particularly well, which had consequences on driving $R_{\star}$ and the inclination $\psi$ to higher values than what we see in our Kepler estimates. This is largely due to the less defined shape of the TESS pulse compared with its Kepler counterpart.

\subsection{Orbit Inclination and Impact Parameter}

KMM18 does not report an orbital inclination in their fitted parameters.  Instead, they fit to a normalized impact parameter denoted by $b\in(0, 1)$.  Expressed in terms of our notation, we find that this is equivalent to

\begin{equation}
\frac{u_0}{(u_0)_{\rm max}} = \frac{\cos(i)}{\cos(i_{\rm min})}= 0.09\, ,
\label{bratio}
\end{equation}
where $i_{\rm min}$ is the minimum required orbital inclination for a self-lensing event at mid-eclipse as found in \citet{2021MNRAS.507..374W} given by 
\begin{equation}
i_{\rm min}=\cos^{-1}\left(\frac{R_{\rm E}(t_0)+R_\star}{d(t_0)}\!\right) \, .
\label{imin}
\end{equation}
For KMM18, $i_{min}\approx89.49^{\circ}$ which, by plugging into equation (\ref{bratio}), allows us to find the orbital inclination of $i=89.95^{\circ}$.  We then use error propagation to find the error bars for this value as reported in Table \ref{tab:1}.

\label{lastpage}
\end{document}